\documentclass[twoside]{dis08}
\usepackage[latin1]{inputenc}
\usepackage[dvips]{graphicx,epsfig,color}
\usepackage{wrapfig,rotating}
\usepackage{amssymb,amsmath,array}

\begin{document}
\title{HERA-signature based searches I: events with isolated leptons and missing transverse momentum at HERA}

%***********************************************************************
% AUTHORS INFORMATION AREA
%***********************************************************************
\author{James Ferrando
%
% Optional short acknowledgment: remove next line if non-needed
%\thanks{This is an optional funding source acknowledgment.}
%
% DO NOT MODIFY THE FOLLOWING '\vspace' ARGUMENT
\vspace{.3cm}\\
%
% Addresses and institutions (remove "1- " in case of a single institution)
%1- School of First Author - Dept of First Author \\
%Address of First Author's school - Country of First Author's
%school
University of Oxford - Dept of Physics \\
Denys Wilkinson Building  
Keble Road, Oxford  
OX1 3RH - United Kingdom
%
% Remove the next three lines in case of a single institution
%\vspace{.1cm}\\
%2- School of Second Author - Dept of Second Author \\
%Address of Second Author's school - Country of Second Author's school\\
}
%***********************************************************************
% END OF AUTHORS INFORMATION AREA
%***********************************************************************

\maketitle

\begin{abstract}
Recent results of searches for events containing isolated leptons
and missing transverse momentum at HERA are presented. Searches in this final
state have yielded notable excesses over Standard Model expectations in the past.
Searches for isolated leptons in channels corresponding to all three
generations of leptons over the full HERA running period are now available.
The combined H1+ZEUS results for searches for electrons and muons are 
compatible with the Standard Model.
\end{abstract}

\section{Introduction}
Searches for events containing isolated leptons and large  missing transverse
momentum have long been considered to be a high priority at HERA. The
dominant genuine source of such events within the Standard Model (SM) is real 
$W$-boson  production, with subsequent leptonic decay, there is also a small 
expected event rate  from real $Z$-boson production with subsequent decay to
 neutrinos. The total  expected cross section for these heavy boson production
  processes at HERA is  of the order of $1 \mathrm{pb}$ at leading 
order \cite{Baur:1991pp}. Quantum chromodynamics (QCD) corrections to this 
cross-section at next-to-leading-order (NLO) have been calculated 
\cite{Diener:2002if}: the overall magnitude of the cross section is not 
changed; the uncertainty on the expected $W$-production cross section after
 including these corrections is approximately $15\%$. Mismeasurement  of events produced by other SM processes such as neutral current (NC) deep inelastic scattering (DIS), charged current (CC) DIS and dilepton production also produce measurable event rates matching this topology and such events must be removed with cuts; this class of events are referred to as 'fake signal'.

Events  containing isolated leptons and large  missing transverse
momentum would be produced at HERA by several models of physics beyond the SM such as single-top-production via flavour-changing neutral currents \cite{Fritzsch:1999rd} and $\mathcal{R}$-parity violating supersymmetry \cite{Choi:2006ms}. 
Both SM vector-boson production and these new physics models are expected to produce leptons with large values of transverse momentum ($P_T^l$).
However, the new physics models would be expected to produce events with
large values of hadronic transverse momentum ($P_T^X$) whereas SM vector boson
production is expected to predominantly produce events with small $P_T^X$. 
The low expected production rate of these events from the SM  
and the clean topology make this an ideal environment in which to search  for evidence for these new physics models. 

Previous searches for isolated electrons and muons by the H1 collaboration
using data taken during the 1994-2000 (HERA-I) running period \cite{Adloff:1998aw,Andreev:2003pm} yielded an
excess of events over the SM predictions at large values ($> 25\,\mathrm{GeV}$)
of $P_T^X$. 
Searches for isolated-electron events in the HERA-I data by the ZEUS 
collaboration in the context of single-$W$  and single top production
\cite{Breitweg:1999ie,Chekanov:2003yt} did not confirm this excess. Searches
for isolated-tau events in the HERA-I data were also performed by both  
H1\cite{Aktas:2006fc} 
and ZEUS\cite{Chekanov:2003bf}, the results were compatible with SM predictions.
With a four-fold increase in integrated luminosity relative to HERA-I, the  2003-2007 
(HERA-II) running period offers data that can resolve the apparent 
discrepancy between the previous results of the H1 and ZEUS collaborations.

 In these proceedings the latest results from searches for isolated-electron 
 and isolated-muon events at H1 and ZEUS are shown. The result of combining these H1 and ZEUS  searches is presented. Finally an H1 search for isolated tau leptons is presented.
These new results include data collected during the 
HERA-II running period. Results that interpret searches for  
this class of  events in the context of measurements of single-$W$ 
production or  single-top production are discussed elsewhere \cite{rizvi:proc}.

\section{Searches for isolated electrons or muons}

\begin{wrapfigure}{r}{0.41\columnwidth}
\centerline{\includegraphics[width=0.37\columnwidth]{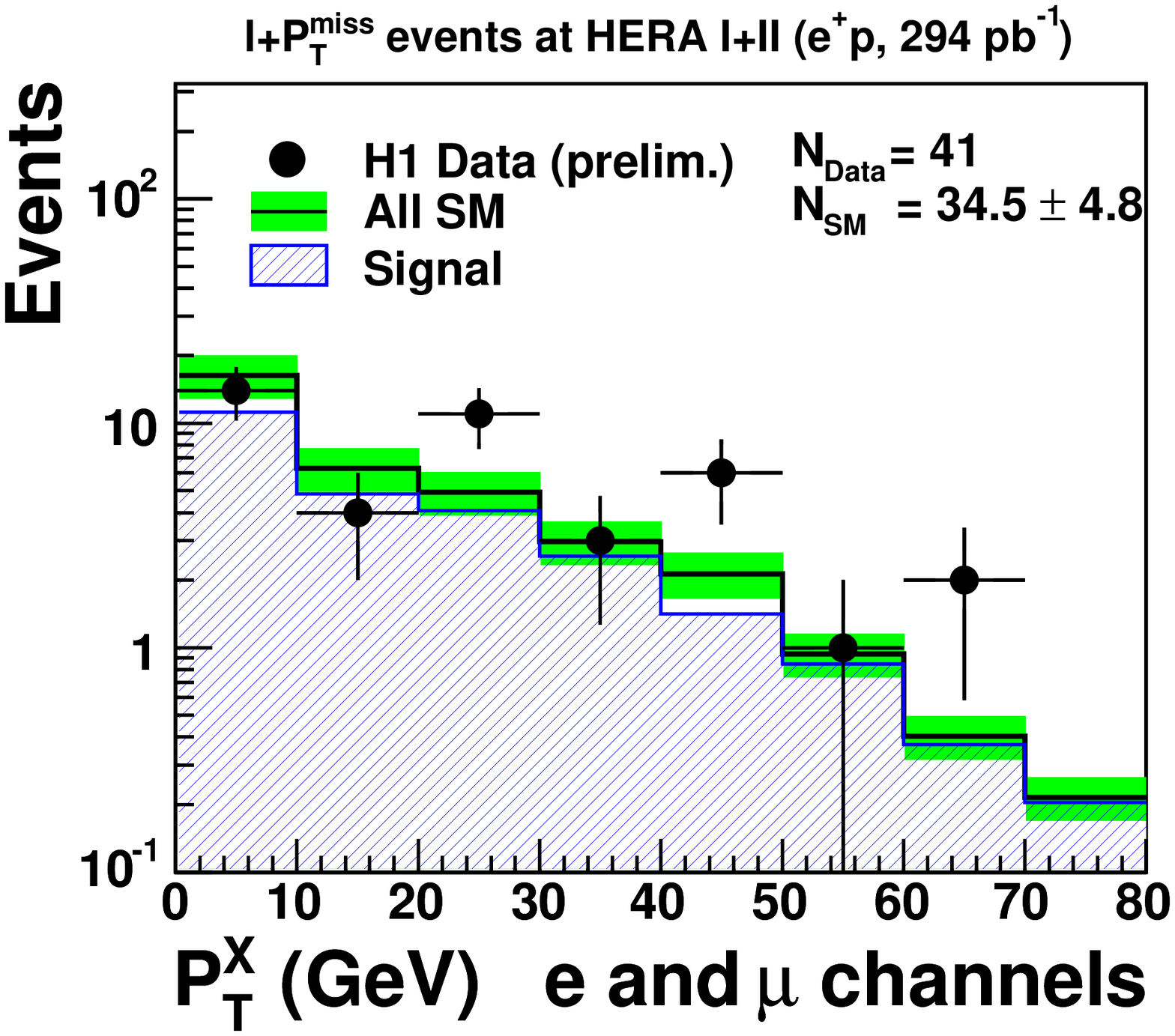}}
\centerline{\includegraphics[width=0.37\columnwidth]{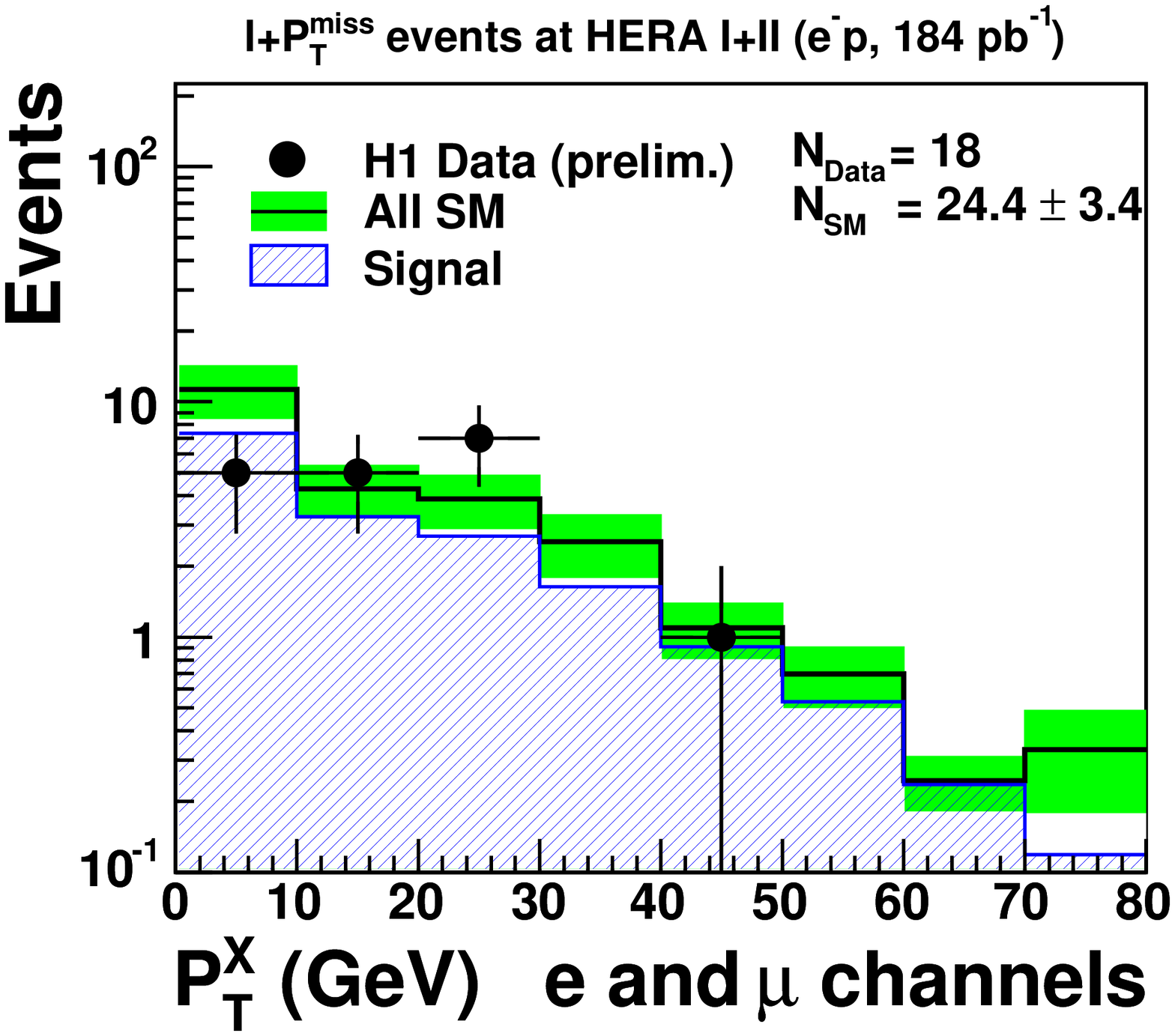}}
\caption{The distribution of $P_T^X$ for events selected in the H1 isolated lepton search for (top) $e+p$ collisions and (bottom) $e^-p$ collisions.}\label{fig:h1lep}
\end{wrapfigure}

Searches for events containing isolated leptons (electrons or muons) and large  missing 
transverse momentum have recently been performed at HERA by both the
H1 \cite{h1prel:07-063} and ZEUS \cite{ZEUSprel:07-021} collaborations. The main
source of fake signal in the electron search are CC DIS at low values of $P_T^X$ and NC DIS at high $P_T^X$. In the muon search the main source of fake signal
is Bethe-Heitler dimuon production. In both searches the fraction of the SM
prediction arising from $W$ production is typically over $60\%$

The H1 search is performed in the same kinematic region as was explored in the previous HERA-I search \cite{Andreev:2003pm}. This region is defined by
$P_T^l>10\,\mathrm{GeV}$ and the polar angle of the lepton, $\theta_l$, lying
in the range $ 5 < \theta_l < 140^{\circ}$. The kinematic region in the muon
phase space is additionally limited to $P_T^X>12\,\mathrm{GeV}$. The data set considered corresponds to an integrated luminosity of $294\,\mathrm{pb^{-1}}$ ($184\,\mathrm{pb^{-1}}$) for $e^+p$ ($e^-p$) collisions. 

 The  distribution of $P_T^X$ is compared to SM Monte Carlo (MC) separately for $e^+p$ and $e^-p$ collision data in Fig. \ref{fig:h1lep}. It can be seen that in  $e^+p$ collsions
the data agree very well with the SM at small values of $P_T^X$ but is generally above the SM predictions at larger values of $P_T^X$. For $P_T^X>25\,\mathrm{GeV}$,  21 (3) events are observed where $8.9\pm1.5$ ($6.9 \pm 1.0$) are expected for $e^+p$ ($e^-p$) collsions. The excess of data over SM in $e^+p$ collisions is present in both the electron and muon searches, where the numbers seen are respectively 11 and 10 compared to expectations of  $4.7\pm 0.9$ and $4.2\pm 0.7$.

\begin{figure}[h]
\centerline{\includegraphics[width=0.45\columnwidth]{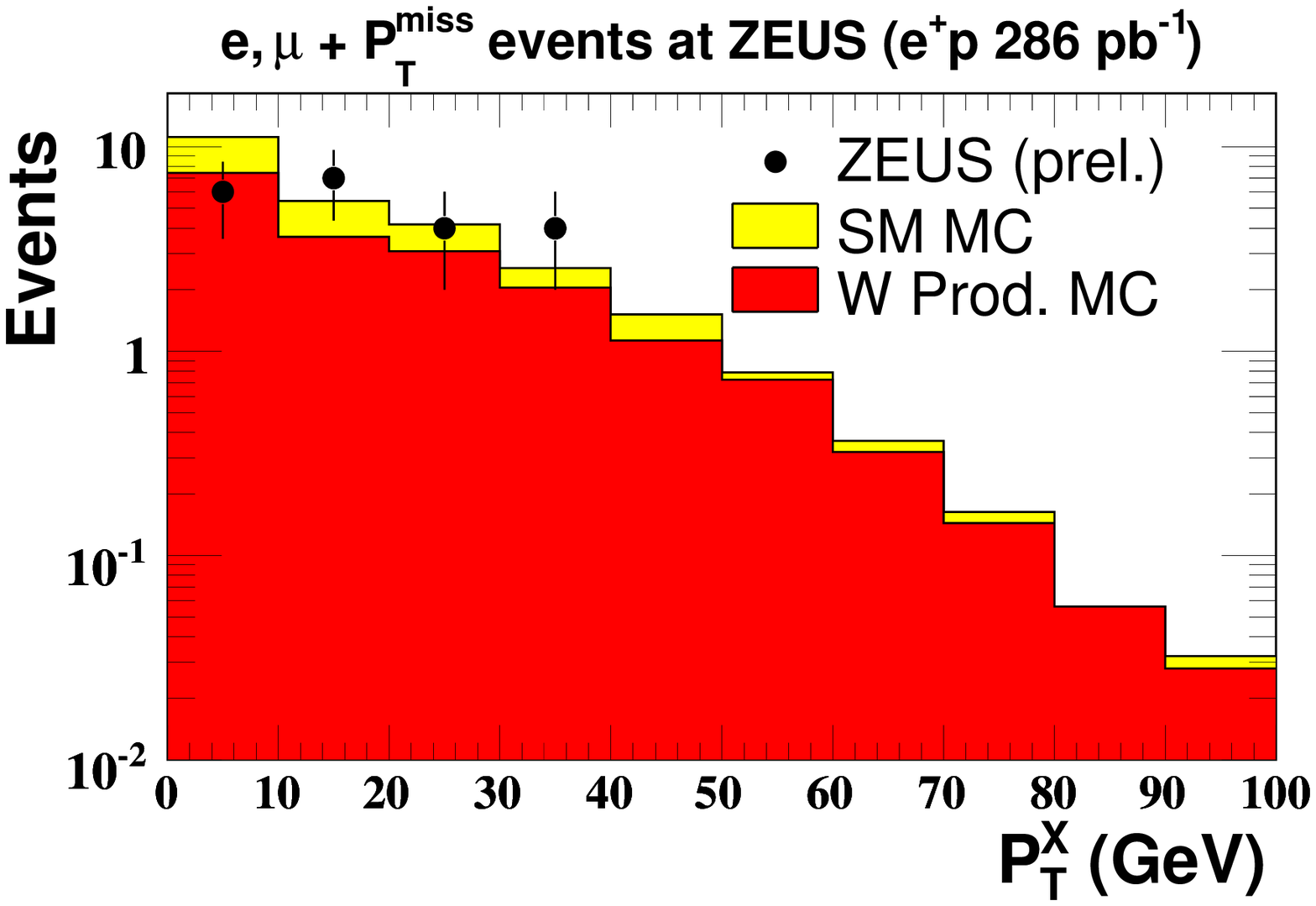} \includegraphics[width=0.45\columnwidth]{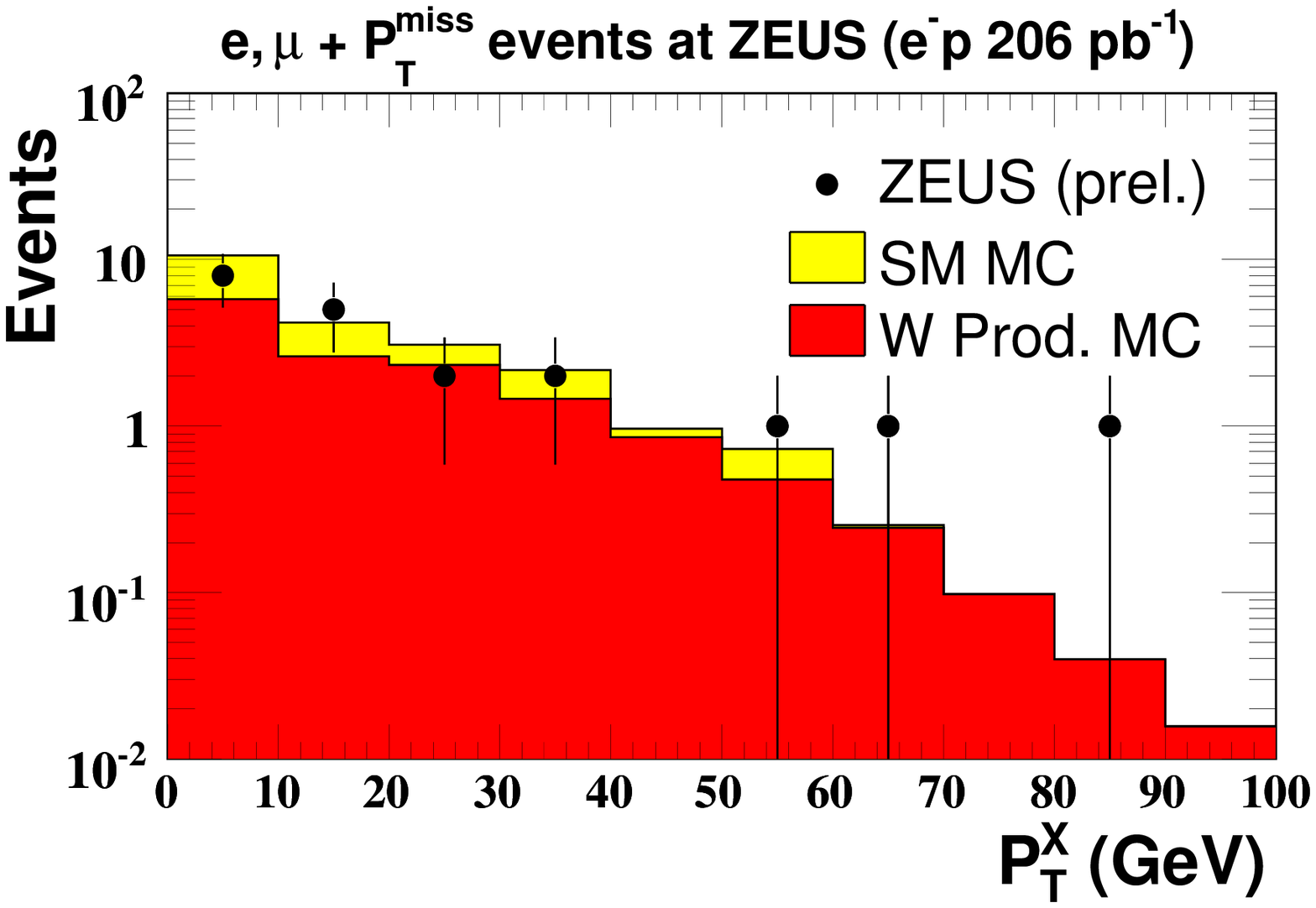}}
\caption{The distribution of $P_T^X$ for events selected in the ZEUS isolated-lepton search for (left) $e+p$ collisions and (right) $e^-p$ collisons.}\label{fig:zeuslep}
\end{figure}

The ZEUS search is performed in a kinematic region defined by
$P_T^l>10\,\mathrm{GeV}$ and  $ 15 < \theta_l < 120^{\circ}$. The kinematic 
region in the muon phase space is additionally limited to $P_T^X>12\,\mathrm{GeV}$.  The data set considered corresponds to an integrated luminosity of $286\,\mathrm{pb^{-1}}$ ($206\,\mathrm{pb^{-1}}$) for $e^+p$ ($e^-p$) collisions. 
 The 
distribution of $P_T^X$ is compared to SM MC separately for $e^+p$ and $e^-p$ 
collision data in Fig. \ref{fig:zeuslep}. It can be seen that the data agrees 
well with the MC at High $P_T^X$. A small deficit of data with respect to MC 
can be seen at low $P_T^X$ in $e^+p$ collisions, similar to that observed by
 the H1 collaboration in $e^-p$ collisions.

\section{Combined H1+ZEUS search}

\begin{wrapfigure}{r}{0.45\columnwidth}
\centerline{\includegraphics[width=0.41\columnwidth]{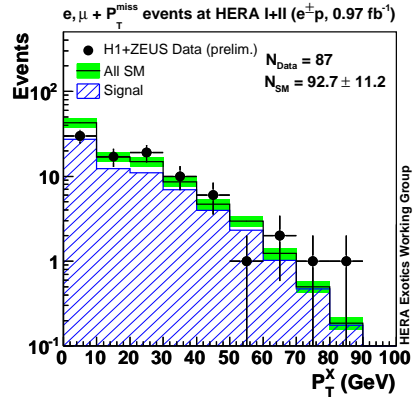}}
\caption{The distribution of $P_T^X$ for the H1+ZEUS combined isolated-lepton search.}\label{fig:h1zeuslep}
\end{wrapfigure}

The H1 and ZEUS isolated-electron or -muon searches are both sensitive to the
same physics processes, with similar efficiencies and fake rates. This enables
the searches to be combined at the data level in the common kinematic region:
$P_T^l>10\,\mathrm{GeV}$ and  $ 15 < \theta_l < 120^{\circ}$. The kinematic 
region in the muon phase space is additionally limited to  $P_T^X>12\,\mathrm{GeV}$. This combination has recently been performed \cite{h1prel:07-162} yielding
a total integrated luminosity of $0.97\,\mathrm{fb^{-1}}$.

 The  distribution of $P_T^X$ is compared to SM MC separately for $e^+p$ and 
$e^-p$  collision data in Fig. \ref{fig:h1zeuslep}. The data agree beautifully with the SM predictions across the full $P_T^X$ range. For $P_T^X>25,\mathrm{GeV}$, 29 events are observed compared to an expectation of $25.3 \pm 3.2$. This 
combined result is compatible with SM $W$ production.

\section{Searches for isolated taus}

\begin{wrapfigure}{l}{0.45\columnwidth}
\centerline{\includegraphics[width=0.41\columnwidth]{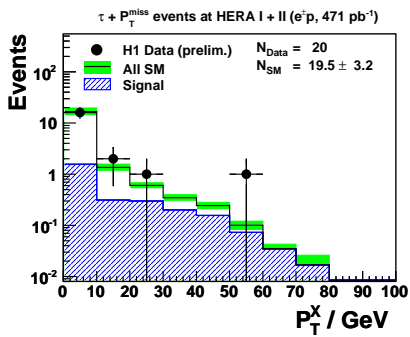}}
\caption{The distribution of $P_T^X$ for events selected in the H1 $\tau$ search.}\label{fig:h1tau}
\end{wrapfigure}

Searches for  events containing isolated $\tau$ leptons and large  missing 
transverse momentum at HERA are extremely challenging. The short lifetime of 
the $\tau$ means that it decays before it can be directly detected. Production
 of isolated taus which decay into leptons 
($\tau^- \rightarrow e^-\bar{\nu}_e \nu_{\tau}$, $\mu^-\bar{\nu}_{\mu} \nu_{\tau} $
 and charge conjugates)  are experimentally indistinguishable from direct
 production of isolated electrons or muons. The $\tau$ decay channel of choice
for these searches at high transverse momentum is therefore the single-prong hadronic decay. Different algorithms  are used to separate the highly collimated 
hadronic jets produced by isolated taus from jets produced by other sources.
Charged current DIS events where $\tau$ candidates are found in the hadronic final state constitute the largest fake signal contribution in this search.

The H1 collaboration has recently produced a new search for isolated $\tau$ 
leptons \cite{h1prel:07-064} using $287\,\mathrm{pb}^{-1}$ ($184\,\mathrm{pb}^{-1}$) data from 
$e^+p$ ($e^-p$) collisions in both HERA-I and HERA-II running. The search is 
performed for events with $P_T^{\tau}>10\,\mathrm{GeV}$ and $20 <  \theta_{\tau} < 120$. The distribution of events found is shown in Fig. \ref{fig:h1tau}, no excess over the SM predictions is  observed in either $e^+p$ or $e^-p$ collisions. For events in the  
$P_T^{X}> 25\,\mathrm{GeV}$ region, 1 data event is observed  compared to an expectation  of $0.99 \pm 0.13$. It should be noted that the fraction of events
arising from $W$ production in this search is significantly smaller than  in
the $e$ and $\mu$ searches.

\section{Summary}

The results of searches for events containing isolated leptons and 
large  missing  transverse momentum using data collected over the entire HERA 
running period have been presented.  The latest electron- and
 muon-search results from H1 reveal no excess over the SM predictions for the 
$e^-p$ data, however the previously observed excess over SM predictions in
 $e^+p $ data persists. The  ZEUS collaboration has also searched for 
isolated electrons and muons, no excess over SM predictions is observed.
In the new $\tau$ search   performed by the H1 collaboration no excess of 
events over the SM is observed. Collaboration between H1 and ZEUS has ensured 
that their most recent electron and muon searches are compatible, enabling
combination at the event level. The results of the combined search for
electrons and muons are compatible with the Standard Model.

% ****************************************************************************
% BIBLIOGRAPHY AREA
% ****************************************************************************

\begin{footnotesize}
% IF YOU DO NOT USE BIBTEX, USE THE FOLLOWING SAMPLE SCHEME FOR THE REFERENCES
% ----------------------------------------------------------------------------
%\begin{thebibliography}{99}
% Please replace the numbers for   contribId   and   sessionId
% in the following URL. You can get this information by going to 
% http://indico.cern.ch/confAuthorIndex.py?confId=24657
% and search for your contribution and click on the title
% Be aware: '&amp;' must be replaced by simple '&' as in example below
% ----------------------------------------------------------------------------

% IF YOU USE BIBTEX,
% - DELETE THE TEXT BETWEEN THE TWO ABOVE DASHED LINES
% - UNCOMMENT THE NEXT TWO LINES AND REPLACE 'Name_Of_Your_BibFile'
%\bibliographystyle{ferrando_james}
%\bibliography{ferrando_james}

%\end{bibliography}

% ****************************************************************************
% END OF BIBLIOGRAPHY AREA
% ****************************************************************************
\end{footnotesize}
\end{document}